\documentclass{article}
\usepackage{amssymb}

\usepackage{amsmath}

\input{tcilatex}

\begin{document}

\title{Extra Dimensions and Nonlinear Equations}
\author{{\Large Thomas Curtright}$\,^{\S }$ and {\Large David Fairlie}$%
\,^{\boxtimes }\medskip $ \\
\textit{Curtright@physics.miami.edu}\ \ \ \ \ \textit{%
David.Fairlie@durham.ac.uk}\\
$^{\S }\,$Department of Physics, University of Miami, Coral Gables, Florida
33124-8046, USA\\
$^{\boxtimes }\,$Department of Mathematical Sciences, University of Durham,
Durham, DH1 3LE, UK}
\date{}
\maketitle

\begin{abstract}
Solutions of nonlinear multi-component Euler-Monge partial differential
equations are constructed in $n$ spatial dimensions by \emph{%
dimension-doubling}, a method that completely linearizes the problem. \
Nonlocal structures are an essential feature of the method. \ The
Euler-Monge equations may be interpreted as a boundary theory arising from a
linearized bulk system such that all boundary solutions follow from simple
limits of those for the bulk. \ 
\end{abstract}

For any theory\footnote{%
We dedicate this paper to Peter Freund on the occasion of his becoming
Professor Emeritus at the University of Chicago, and thereby begin with
general remarks about the origin of extra dimensions, allowing for the
possibility that these are similar to but not necessarily on the exact same
footing as the original dimensions. \ For related points of view within the
Kaluza-Klein physics framework \cite{Appelquist}, see \cite{Arkani-Hamed}.}
with an infinite number of conservation laws, we may always assemble the
conserved currents into a generating function involving a spectral parameter 
$a$. \ If that spectral parameter is independent of any other spacetime
dimensions in the theory, as is possible in the simplest cases, then
effectively the theory possesses an \emph{extra dimension}\footnote{%
More precisely, an extra \emph{bosonic} dimension. \emph{\ }A \emph{finite}
number of conservation laws evokes extra \emph{fermionic} or \emph{anyonic
dimensions}, $\theta $, involving $k$th order superspace or Grassman
variables. \ This and other non-commutative geometries will not be discussed
further here.}. \ Moreover, it is always possible to openly include this
extra dimension in some of the dynamical equations, and not just leave it as
a \emph{dimension sub rosa}. \ 

For example, suppose a theory is originally expressed in terms of
coordinates $\left( x,t\right) $ with an infinite number of conserved
currents: $\partial _{t}\rho ^{\left( n\right) }\left( x,t\right) =\partial
_{x}J^{\left( n\right) }\left( x,t\right) $, $n\in \mathbb{N}$. \ Then by
defining $\rho \left( x,t,a\right) \equiv \sum_{n}\left( n+1\right)
a^{n}\rho ^{\left( n\right) }\left( x,t\right) $, as opposed to $%
\sum_{n}a^{n}\rho ^{\left( n\right) }\left( x,t\right) $, and $J\left(
x,t,a\right) \equiv \sum_{n}a^{n+1}J^{\left( n\right) }\left( x,t\right) $,
as opposed to $\sum_{n}a^{n}J^{\left( n\right) }\left( x,t\right) $, we have
rendered all the conservation laws as a single second-order
higher-dimensional partial differential equation (PDE): $\partial _{t}\rho
\left( x,t,a\right) =\partial _{x}\partial _{a}J\left( x,t,a\right) $, as
opposed to the first-order $\partial _{t}\rho \left( x,t,a\right) =\partial
_{x}J\left( x,t,a\right) $. \ Hence our choice for the current generating
functions has fully exposed an extra dimension in the PDEs satisfied by
those generating functions. \ The extra dimension here does not just ride
along as a suppressible label for the currents but it appears explicitly,
perhaps even unavoidably, in the dynamical equations. \ Of course this
immediately raises issues about whether the theory requires $a$ to appear
explicitly for \emph{all} dynamical equations to be cogently expressed in
terms of the original plus extra dimensions, and about covariance properties
for the theory in the complete set of dimensions. \ 

In this paper we address these issues for a simple but very generally
applicable class of nonlinear PDE's \cite{Euler,Monge}: \ The first order
Euler-Monge (E-M) equations\ $\partial _{t}\mathbf{u}=\left( \mathbf{u}\cdot
\nabla \right) \mathbf{u}$. \ We find the full dynamics of these \emph{%
nonlinear} theories are elegantly encoded into a higher dimensional set of 
\emph{linear} ``heat'' equations obtained through dimension doubling $\left( 
\mathbf{x}\right) \rightarrow \left( \mathbf{x},\mathbf{a}\right) $, where
for each spatial coordinate $x_{i}$ there is an \emph{associated coordinate}
given by spectral parameter $a_{i}$. \ The original dynamical variables are
obtained as spectral parameter boundary limits, $U_{i}\left( \mathbf{x},t,%
\mathbf{a}\right) \,_{\overrightarrow{\;\mathbf{a\rightarrow 0\;}}%
}\,u_{i}\left( \mathbf{x},t\right) $. \ The fact that the higher dimensional
theory is linearized strongly argues that this is the right approach to
take. \ In the linearized theory, the pairs $\left( x_{i},a_{i}\right) $ act
like ``light-cone'' variables in the enlarged set of dimensions such that
the heat equations for all the dynamical variables are of the form $\left(
\partial /\partial t-\sum_{j=1}^{n}\partial ^{2}/\partial a_{j}\partial
x_{j}\right) U_{i}\left( \mathbf{x},t,\mathbf{a}\right) =0$. \ Thus the
extra dimensions appear explicitly and, indeed, unavoidably in these
linearized dynamical equations.

We also find Nambu brackets \cite{Nambu} of the fields, of all orders up to
the full Jacobian, as a remarkable feature of the linearizing maps. \ We
know of only one other field theoretic example \cite{Baker} where these
brackets appear so naturally. \ Moreover, the linearizing maps are \emph{%
nonlocal} in all but the simplest, one component case. \ The nonlocal
structures appropriate for E-M equations with two components in two spatial
dimensions are evocative of phase factors in Wilson loops (cf. strings), and
when the E-M equations describe $n$ component fields in the original $n$
spatial dimensions these structures extend to higher dimensional
constructions involving integrals over $n-1$ dimensional submanifolds (cf. ($%
n-1$)-branes). In the one dimensional, one component case, the E-M solution
is obtained algebraically from the dimensionally-doubled ``bulk'' solution
for all values of the single spectral parameter. \ In higher dimensional or
multi-component cases the dependence of the solutions on the spectral
parameters is more involved. \ Nevertheless, in all cases the solutions of
the E-M equations may be obtained from simple limits of those for the bulk.

The Euler-Monge equations first appeared in 18th and 19th century studies of
fluid dynamics \cite{Euler} and analytic geometry \cite{Monge}. \ Riemann
took up a study of the equations in the context of gas dynamics, discussing
the equations as a theory of invariants \cite{Riemann}\ (for a modern
textbook treatment, see \cite{Debnath}). \ His approach is widely applicable
to almost all nonlinear flow problems, although it does not triumph over
turbulence. \ A systematic modern discussion of the E-M equations that
synthesizes ideas from both geometry and invariance theory can be found in
the review by Dubrovin and Novikov \cite{Dubrovin}. \ Most contemporary
texts and reviews stress the universal role played by these nonlinear
transport equations in accordance with Whitham's theory \cite{Whitham}. \
Essentially all nonlinear waves, even those in dispersive and dissipative
media, involve E-M equations, or simple variants of them, if the nonlinear
wavetrains are slowly varying. \ This makes the equations particularly
useful for analyzing the asymptotic behavior of nonlinear solutions. \ The
E-M equations and their conservation laws also serve as a useful starting
point in Polyakov's study of turbulence \cite{Polyakov} but without yet
leading to a general solution of the Navier-Stokes equations.

The first order E-M equation $\frac{\partial u}{\partial t}=u\frac{\partial u%
}{\partial x}$ also gives rise to the Bateman equation \cite{Bateman} upon
substituting $u=\frac{\partial \phi /\partial t}{\partial \phi /\partial x}$%
. \ The resulting second order nonlinear PDE is $\phi _{x}^{2}\phi
_{tt}-2\phi _{x}\phi _{t}\phi _{tx}+\phi _{t}^{2}\phi _{xx}=0$, and is well
known to possess a general implicit solution given by solving $tS_{0}(\phi
)+xS_{1}(\phi )=\mathrm{constant}$, where $S_{0}$ and $S_{1}$ are arbitrary
differentiable functions of $\phi (x,t)$. \ The structure of this solution
incorporates the covariance properties of the PDE: \ If $\phi $ is a
solution, so is any function of $\phi $. \ In fact, curiously, the
generalization of this solution to $n+1$ functions $S_{0}(\phi ),S_{i}(\phi
) $ of $\phi \left( \mathbf{x},t\right) $, $\mathbf{x=}\left( x_{1},\cdots
,x_{n}\right) $, subject to a single constraint $tS_{0}(\phi )+\sum
x_{i}\,S_{i}(\phi )=0$, is a ``universal solution'' \cite{dbf4} to \emph{any}
equation derived from a Lagrangian which is homogeneous of weight one in the
first derivatives of $\phi $.

Thus the Euler-Monge equations appear widespread across a very broad
landscape of physics and applied mathematics problems, and therefore it is
important to understand their solutions at as many levels as possible. \ To
that end we shall map all solutions of the E-M equations in arbitrary
dimensions into solutions of second-order linear equations. \ This type of
map is reminiscent of the Cole-Hopf \cite{ColeHopf,Forsyth} transformation
(thoroughly reviewed in \cite{Gesztesy}) used to linearize the Burgers \cite%
{Burgers,Forsyth} nonlinear diffusion equation, but there are important
differences here. \ The Cole-Hopf transformation only works for curl-free $%
\mathbf{u}$, does not use extra dimensions, and fails for $0=\kappa $ (the
diffusivity). \ The map to follow works for all $\mathbf{u}$, curl-free or
otherwise, does use extra dimensions, but works only for $\kappa =0$. \ (We
hope to extend the method to $\kappa \neq 0$ in a subsequent study.)

We believe it is most efficient to present our results summarily as a small
set of Theorems, for which we sketch the salient features of their proofs. \
In the following, $\mathcal{M}_{n}$ is the $n$ dimensional nonlinear
Euler-Monge operator and $\mathcal{H}_{n}$ is an associated hyperbolic heat
operator (introduced in \cite{Polyakov}).%
\begin{equation*}
\mathcal{M}_{n}\equiv \frac{\partial }{\partial t}-\sum_{j=1}^{n}u_{j}\frac{%
\partial }{\partial x_{j}}\;,\;\;\;\;\;\mathcal{H}_{n}\equiv \frac{\partial 
}{\partial t}-\sum_{j=1}^{n}\frac{\partial ^{2}}{\partial x_{j}\partial a_{j}%
}
\end{equation*}%
To begin, however, we will generalize these two definitions to allow for an
arbitrary function $F$ in the most elementary results in one spatial
dimension. \ We find\footnote{%
We enumerate the Theorems by ``\emph{k} in \emph{d}'', where \emph{k}\ is
the number of field components and \emph{d}\ is the number of spatial
dimensions in the Euler-Monge equations.}

\paragraph{Theorem 1 in 1:}

$\frac{\partial }{\partial t}U\left( x,t,a\right) =F\left( \frac{\partial }{%
\partial a}\right) \frac{\partial }{\partial x}U\left( x,t,a\right) $\ \ if
and only if $\ \frac{\partial }{\partial t}u\left( x,t\right) =F\left(
u\left( x,t\right) \right) \frac{\partial }{\partial x}u\left( x,t\right) $
\ where%
\begin{equation*}
U\left( x,t,a\right) \equiv \frac{e^{au\left( x,t\right) }-1}{a}%
\;,\;\;\;u\left( x,t\right) =\frac{1}{a}\ln \left( 1+aU\left( x,t,a\right)
\right)
\end{equation*}%
and$\ F$\ is any function with a formal power series. \ \smallskip \newline
\textbf{Proof of 1 in 1:} \ By direct calculation 
\begin{equation*}
\left( \frac{\partial }{\partial t}-F\left( \frac{\partial }{\partial a}%
\right) \frac{\partial }{\partial x}\right) \frac{e^{au\left( x,t\right) }-1%
}{a}=e^{au\left( x,t\right) }\left( \frac{\partial }{\partial t}u\left(
x,t\right) -F\left( u\left( x,t\right) \right) \frac{\partial }{\partial x}%
u\left( x,t\right) \right)
\end{equation*}%
and the Theorem follows. \ $\blacksquare $

\paragraph{Corollary 1 in 1:}

The formal solution for $U\left( x,t,a\right) $ \ in terms of $U\left(
x,t=0,a\right) $ \ is given by%
\begin{equation*}
\left( e^{au\left( x,t\right) }-1\right) /a=e^{t\,F\left( \frac{\partial }{%
\partial a}\right) \frac{\partial }{\partial x}}\left( \left( e^{au\left(
x\right) }-1\right) /a\right)
\end{equation*}%
with $u\left( x,t=0\right) =u\left( x\right) $ . \ This is an elementary
consequence of the Theorem. \ $\blacksquare $

The bulk solution $U\left( x,t,a\right) $ may also be viewed as a simple
one-parameter deformation of the boundary data $u\left( x,t\right) $, with
the extra dimension serving as the deformation parameter. \ In this
exceptional one-component case, we may easily extract $u\left( x,t\right) $
from $U\left( x,t,a\right) $ for any value of the extra dimension $a$ as
given by the logarithmic expression in the Theorem. \ But in particular, we
may extract $u\left( x,t\right) $ as a limit of the bulk solution $u\left(
x,t\right) =\lim\limits_{a\rightarrow 0}U\left( x,t,a\right) $. \ This
immediately yields the time series solution \cite{dbf3} to the previous E-M
equation as$\ $the limit 
\begin{equation*}
u\left( x,t\right) =\lim\limits_{a\rightarrow 0}e^{t\,F\left( \frac{\partial 
}{\partial a}\right) \frac{\partial }{\partial x}}\left( \frac{e^{au\left(
x\right) }-1}{a}\right) =F^{-1}\left[ \sum\limits_{j=0}^{\infty }\frac{t^{j}%
}{\left( 1+j\right) !}\frac{d^{j}}{dx^{j}}\left( F\left[ u\left( x\right) %
\right] \right) ^{1+j}\right]
\end{equation*}%
where we assume $F$ (locally) invertible in the last step\footnote{%
Given the close relation of the Monge and Bateman equations, it might be
expected that the latter also admits a power series solution of simple form.
\ Indeed this is so. \ Treating the equation as hyperbolic with initial
conditions $\phi (x,0)=f(x),\ \partial \phi (x,0)/\partial t=g(x)$, the time
series solution of the Bateman equation is 
\begin{equation*}
\phi (x,t)=f(x)+tg(x)+\sum_{j=1}^{\infty }\frac{t^{1+j}}{(1+j)!}\frac{d^{j}}{%
dx^{j}}\left( \frac{g(x)^{1+j}}{\left( df(x)/dx\right) ^{j}}\right)
\end{equation*}%
}. \ Similar Corollaries and time series solutions are obvious consequences
of all our results, and may be incorporated directly into each Theorem. \
For example, one independent field $u$ in spatial dimensions $\left(
x,y_{1},\cdots ,y_{n}\right) $ with dependent ``velocity fields'' $\left(
u,v_{1}\left( u\right) ,\cdots ,v_{n}\left( u\right) \right) $ leads to

\paragraph{Theorem 1 in (n+1):}

$\frac{\partial }{\partial t}u\left( x,\mathbf{y},t\right) =u\left( x,%
\mathbf{y},t\right) \frac{\partial }{\partial x}u\left( x,\mathbf{y}%
,t\right) +\sum_{i=1}^{n}v_{i}\left( u\left( x,\mathbf{y},t\right) \right) 
\frac{\partial }{\partial y_{i}}u\left( x,\mathbf{y},t\right) $\ \ if and
only if%
\begin{equation*}
\int_{0}^{u\left( x,\mathbf{y},t\right) }du\exp \left(
au+\sum_{i=1}^{n}b_{i}v_{i}\left( u\right) \right) =e^{t\left( \frac{%
\partial ^{2}}{\partial x\partial a}+\sum_{i=1}^{n}\frac{\partial ^{2}}{%
\partial y_{i}\partial b_{i}}\right) }\int_{0}^{u\left( x,\mathbf{y}\right)
}du\exp \left( au+\sum_{i=1}^{n}b_{i}v_{i}\left( u\right) \right)
\end{equation*}

\paragraph{Proof of 1 in (n+1): \ }

By direct calculation, with $U\left( x,\mathbf{y},t,a,\mathbf{b}\right)
\equiv \int_{0}^{u\left( x,\mathbf{y},t\right) }du\exp \left(
au+\sum_{i=1}^{n}b_{i}v_{i}\left( u\right) \right) $ ,%
\begin{multline*}
\left( \frac{\partial }{\partial t}-\frac{\partial ^{2}}{\partial x\partial a%
}-\sum_{i=1}^{n}\frac{\partial ^{2}}{\partial y_{i}\partial b_{i}}\right)
U\left( x,\mathbf{y},t,a,\mathbf{b}\right) = \\
\left( \frac{\partial }{\partial t}u\left( x,\mathbf{y},t\right) -u\left( x,%
\mathbf{y},t\right) \frac{\partial }{\partial x}u\left( x,\mathbf{y}%
,t\right) -\sum_{i=1}^{n}v_{i}\left( u\left( x,\mathbf{y},t\right) \right) 
\frac{\partial }{\partial y_{i}}u\left( x,\mathbf{y},t\right) \right) \exp
\left( au\left( x,\mathbf{y},t\right) +\sum_{i=1}^{n}b_{i}v_{i}\left(
u\left( x,\mathbf{y},t\right) \right) \right)
\end{multline*}%
So, as given, the higher dimensional heat equation is satisfied by the
integral form $U\left( x,\mathbf{y},t,a,\mathbf{b}\right) $ if and only if
the given one-component generalization of the E-M equations holds. \ The RHS
of the relation in the Theorem is then just the formal solution of the heat
equation, as in the previous Corollary\footnote{%
As is true for the Bateman equation and the one-component Monge equation in
one spatial dimension, there is a corresponding second order equation for
the \textbf{1 in (n+1)} case which our solution satisfies. \ It is the
so-called ``Universal Field Equation'' which may be obtained by elimination
of $u$ from the first order equations \cite{dbf1,dbf2}.
\par
{}}. \ $\blacksquare $

The last result does not allow for a simple extraction of $u\left( x,\mathbf{%
y},t\right) $ from the integral form of $U\left( x,\mathbf{y},t,a,\mathbf{b}%
\right) $ for non-vanishing $a,\mathbf{b}$. \ However, it does have the
simple limit $\lim\limits_{a,\mathbf{b}\rightarrow 0}U\left( x,\mathbf{y}%
,t,a,\mathbf{b}\right) =u\left( x,\mathbf{y},t\right) $, so extraction is
trivial on the boundary $a,\mathbf{b}\rightarrow 0$. \ This is true of all
the heat equation solutions to follow. \ Also note, $U\left( x,\mathbf{y}%
,t,a,\mathbf{b}\right) $ in this one-component case is an integral over the
field value. Nevertheless $U$ is still local in all the dimensions, no
matter how many. \ Locality in the original spatial dimensions will \emph{not%
} hold, however, for maps of multi-component fields in higher dimensions. \
This is first illustrated by

\paragraph{Theorem 2 in 2:}

$\mathcal{H}_{2}U=\mathcal{H}_{2}V=0$ \ if and only if \ $\mathcal{M}_{2}u=%
\mathcal{M}_{2}v=0$ \ where ($\;\varepsilon \left( s\right) \equiv \pm \frac{%
1}{2}$ for $s\gtrless 0$\ )%
\begin{eqnarray*}
U\left( x,y,t,a,b\right) &\equiv &\int_{-\infty }^{\infty }dr\;\varepsilon
\left( y-r\right) \;e^{au\left( x,r,t\right) +bv\left( x,r,t\right) }\;\frac{%
\partial u\left( x,r,t\right) }{\partial r} \\
V\left( x,y,t,a,b\right) &\equiv &\int_{-\infty }^{\infty }dq\;\varepsilon
\left( x-q\right) \;e^{au\left( q,y,t\right) +bv\left( q,y,t\right) }\;\frac{%
\partial v\left( q,y,t\right) }{\partial q}
\end{eqnarray*}

\paragraph{Proof of 2 in 2: \ }

Again by direct calculation, assuming $u$, $v$, and their derivatives vanish
asymptotically in $x,y$,%
\begin{eqnarray*}
\mathcal{H}_{2}U\left( x,y,t,a,b\right) &=&e^{au\left( x,y,t\right)
+bv\left( x,y,t\right) }\mathcal{M}_{2}u\left( x,y,t\right) \\
&&+b\int_{-\infty }^{\infty }dr\;\varepsilon \left( y-r\right) \;e^{au\left(
x,r,t\right) +bv\left( x,r,t\right) }\left( \frac{\partial u\left(
x,r,t\right) }{\partial r}\mathcal{M}_{2}v\left( x,r,t\right) -\frac{%
\partial v\left( x,r,t\right) }{\partial r}\mathcal{M}_{2}u\left(
x,r,t\right) \right)
\end{eqnarray*}%
\begin{eqnarray*}
\mathcal{H}_{2}V\left( x,y,t,a,b\right) &=&e^{au\left( x,y,t\right)
+bv\left( x,y,t\right) }\mathcal{M}_{2}v\left( x,y,t\right) \\
&&+a\int_{-\infty }^{\infty }dq\;\varepsilon \left( x-q\right) \;e^{au\left(
q,y,t\right) +bv\left( q,y,t\right) }\left( \frac{\partial v\left(
q,y,t\right) }{\partial q}\mathcal{M}_{2}u\left( q,y,t\right) -\frac{%
\partial u\left( q,y,t\right) }{\partial q}\mathcal{M}_{2}v\left(
q,y,t\right) \right)
\end{eqnarray*}%
The complete Theorem then follows by also using the obvious limits $%
\lim\limits_{a,b\rightarrow 0}\mathcal{H}_{2}U\left( x,y,t,a,b\right) =%
\mathcal{M}_{2}u\left( x,y,t\right) $ and $\lim\limits_{a,b\rightarrow 0}%
\mathcal{H}_{2}V\left( x,y,t,a,b\right) =\mathcal{M}_{2}v\left( x,y,t\right) 
$. \ $\blacksquare $

As advertised, the two-component map in two spatial dimensions involves a
nonlocal transformation between E-M and heat equation solutions: \ It
features line integrals over the original spatial variables. \ The map is
still local in the extra dimensions, however. \ This nonlocality in the
original dimensions persists and is even extended when more components and
more spatial dimensions are considered. \ As a further illustration before
giving the generalization to an arbitrary number of dimensions, we have

\paragraph{Theorem 3 in 3:}

$\mathcal{H}_{3}U=\mathcal{H}_{3}V=\mathcal{H}_{3}W=0$ \ if and only if \ $%
\mathcal{M}_{3}u=\mathcal{M}_{3}v=\mathcal{M}_{3}w=0$ \ where%
\begin{eqnarray*}
U\left( x,y,z,t,a,b,c\right) &\equiv &\int dr\,\varepsilon \left( y-r\right)
\,e^{au+bv+cw}\;\frac{\partial u\left( x,r,z,t\right) }{\partial r}-c\iint
drds\,\varepsilon \left( y-r\right) \,\varepsilon \left( z-s\right)
\,e^{au+bv+cw}\left\{ u,w\right\} _{rs}\left( x,r,s,t\right) \\
V\left( x,y,z,t,a,b,c\right) &\equiv &\int ds\,\varepsilon \left( z-s\right)
\,e^{au+bv+cw}\;\frac{\partial v\left( x,y,s,t\right) }{\partial s}-a\iint
dqds\,\varepsilon \left( x-q\right) \,\varepsilon \left( z-s\right)
\,e^{au+bv+cw}\left\{ v,u\right\} _{sq}\left( q,y,s,t\right) \\
W\left( x,y,z,t,a,b,c\right) &\equiv &\int dq\,\varepsilon \left( x-q\right)
\,e^{au+bv+cw}\;\frac{\partial w\left( q,y,z,t\right) }{\partial q}-b\iint
dqdr\,\varepsilon \left( x-q\right) \,\varepsilon \left( y-r\right)
\,e^{au+bv+cw}\left\{ w,v\right\} _{qr}\left( q,r,z,t\right)
\end{eqnarray*}

\paragraph{Proof of 3 in 3: \ }

There are a few essential new ingredients needed to complete the proof by
direct calculation in this case. \ Define Poisson brackets as usual by 
\begin{equation*}
\left\{ u,v\right\} _{rs}=\frac{\partial u}{\partial r}\frac{\partial v}{%
\partial s}-\frac{\partial u}{\partial s}\frac{\partial v}{\partial r}
\end{equation*}%
where $u$ and $v$ are any two functions of the independent variables $r$ and 
$s$. \ Then it is straightforward to show%
\begin{eqnarray*}
\frac{\partial }{\partial t}\left\{ u,v\right\} _{zy}-\frac{\partial }{%
\partial x}\left( u\left\{ u,v\right\} _{zy}\right) -\frac{\partial }{%
\partial y}\left( v\left\{ u,v\right\} _{zy}\right) -\frac{\partial }{%
\partial z}\left( w\left\{ u,v\right\} _{zy}\right) &=&\left\{ \mathcal{M}%
_{3}u,v\right\} _{zy}+\left\{ u,\mathcal{M}_{3}v\right\} _{zy} \\
\frac{\partial }{\partial t}\left\{ u,v\right\} _{xy}-\frac{\partial }{%
\partial x}\left( u\left\{ u,v\right\} _{xy}\right) -\frac{\partial }{%
\partial y}\left( v\left\{ u,v\right\} _{xy}\right) -\frac{\partial }{%
\partial z}\left( w\left\{ u,v\right\} _{xy}\right) &=&\left\{ \mathcal{M}%
u,v\right\} _{xy}+\left\{ u,\mathcal{M}v\right\} _{xy}-\left\{ u,v,w\right\}
_{xyz}
\end{eqnarray*}%
as well as similar relations obtained by permutation of dependent and
independent variables. \ In the last relation we have introduced the totally
antisymmetric Nambu triple bracket (i.e. Jacobian, in this 3-dimensional
case)%
\begin{equation*}
\left\{ u,v,w\right\} _{xyz}=\frac{\partial u}{\partial x}\left\{
v,w\right\} _{yz}+\frac{\partial u}{\partial y}\left\{ v,w\right\} _{zx}+%
\frac{\partial u}{\partial z}\left\{ v,w\right\} _{xy}=\frac{\partial u}{%
\partial x}\left\{ v,w\right\} _{yz}+\frac{\partial v}{\partial x}\left\{
w,u\right\} _{yz}+\frac{\partial w}{\partial x}\left\{ u,v\right\} _{yz}
\end{equation*}%
Once equipped with such relations, the complete proof of the Theorem is
tedious, perhaps, but not subtle. \ (See the generalization to follow for
additional details.) $\ \blacksquare $

The nonlocality appearing in our map for three components in three spatial
dimensions is two-dimensional: \ It features surface integrals over pairs of
the original spatial dimensions, perhaps evocative of membrane-based phase
factors. \ Nonetheless, the map is still local in the extra dimensions and
the E-M solutions are again trivially given by boundary limits of the bulk
constructions. \ The nonlocality is extended to $\left( n-1\right) $%
-dimensional integrals when n-component linearizing maps are constructed in $%
n$ spatial dimensions. \ This is explicit in

\paragraph{Theorem {\protect\large n} in {\protect\large n}:}

$\mathcal{H}_{n}U_{k}\left( \mathbf{x},t,\mathbf{a}\right) =0$ \ if and only
if \ $\mathcal{M}_{n}u_{i}\left( \mathbf{x},t\right) =0$ \ for \ $i,k\in
\left\{ 1,\cdots ,n\right\} $ \ where%
\begin{eqnarray*}
U_{k}\left( \mathbf{x},t,\mathbf{a}\right) &\equiv &\int \cdots \int
dq_{1}\cdots dq_{n}\,\delta \left( q_{k}-x_{k}\right) \left( \frac{%
e^{a_{k}u_{k}}-1}{a_{k}}\right) \times \\
&& \\
&&\times \det \left( 
\begin{array}{ccc}
\frac{\partial }{\partial q_{1}}\left( \varepsilon \left( q_{1}-x_{1}\right)
e^{a_{1}u_{1}}\right) & \cdots & \frac{\partial }{\partial q_{n}}\left(
\varepsilon \left( q_{1}-x_{1}\right) e^{a_{1}u_{1}}\right) \\ 
\vdots & \ddots & \vdots \\ 
\frac{\partial }{\partial q_{1}}\left( \varepsilon \left( q_{n}-x_{n}\right)
e^{a_{n}u_{n}}\right) & \cdots & \frac{\partial }{\partial q_{n}}\left(
\varepsilon \left( q_{n}-x_{n}\right) e^{a_{n}u_{n}}\right)%
\end{array}%
\right) _{\substack{ \text{exclude }k\text{th row}  \\ \text{and }k\text{th
column}}}
\end{eqnarray*}

\paragraph{Proof of {\protect\large n} in {\protect\large n}:%
\protect\footnote{%
After obtaining this result, we learned that Polyakov had found
inhomogeneous versions of the same higher-dimensional ``heat'' equations are
obeyed by certain correlation functions in his statistical treatment of
turbulence \cite{Polyakov}\ (especially p 6188).} \ }

Consider only the first component (et sic de similibus).%
\begin{multline*}
U_{1}\left( \mathbf{x},t,\mathbf{a}\right) =\int \cdots \int dq_{1}\cdots
dq_{n}\,\delta \left( q_{1}-x_{1}\right) \left( \frac{e^{a_{1}u_{1}}-1}{a_{1}%
}\right) \det \left( 
\begin{array}{ccc}
\frac{\partial }{\partial q_{2}}\left( \varepsilon \left( q_{2}-x_{2}\right)
e^{a_{2}u_{2}}\right) & \cdots & \frac{\partial }{\partial q_{n}}\left(
\varepsilon \left( q_{2}-x_{2}\right) e^{a_{2}u_{2}}\right) \\ 
\vdots & \ddots & \vdots \\ 
\frac{\partial }{\partial q_{2}}\left( \varepsilon \left( q_{n}-x_{n}\right)
e^{a_{n}u_{n}}\right) & \cdots & \frac{\partial }{\partial q_{n}}\left(
\varepsilon \left( q_{n}-x_{n}\right) e^{a_{n}u_{n}}\right)%
\end{array}%
\right) \\
=\int \cdots \int dq_{2}\cdots dq_{n}\,\left( \frac{e^{a_{1}u_{1}}-1}{a_{1}}%
\right) \det \left( 
\begin{array}{ccc}
\frac{\partial }{\partial q_{2}}\left( \varepsilon \left( q_{2}-x_{2}\right)
e^{a_{2}u_{2}}\right) & \cdots & \frac{\partial }{\partial q_{n}}\left(
\varepsilon \left( q_{2}-x_{2}\right) e^{a_{2}u_{2}}\right) \\ 
\vdots & \ddots & \vdots \\ 
\frac{\partial }{\partial q_{2}}\left( \varepsilon \left( q_{n}-x_{n}\right)
e^{a_{n}u_{n}}\right) & \cdots & \frac{\partial }{\partial q_{n}}\left(
\varepsilon \left( q_{n}-x_{n}\right) e^{a_{n}u_{n}}\right)%
\end{array}%
\right) \left( x_{1},q_{2},\cdots ,q_{n},t\right) \\
=\int \cdots \int dq_{2}\cdots dq_{n}\,\left( \frac{e^{a_{1}u_{1}}-1}{a_{1}}%
\right) \varepsilon _{i_{2}\cdots i_{n}}\frac{\partial }{\partial q_{i_{2}}}%
\left( \varepsilon \left( q_{2}-x_{2}\right) e^{a_{2}u_{2}}\right) \frac{%
\partial }{\partial q_{i_{3}}}\left( \varepsilon \left( q_{3}-x_{3}\right)
e^{a_{3}u_{3}}\right) \cdots \frac{\partial }{\partial q_{i_{n}}}\left(
\varepsilon \left( q_{n}-x_{n}\right) e^{a_{n}u_{n}}\right)
\end{multline*}%
where in the last expression the $i_{k}$ dummy indices, $k\in \left\{
2,\cdots ,n\right\} $, are summed from $2$ to $n$, i.e. $1$ is excluded. \
Now we integrate by parts assuming all fields and their derivatives vanish
as $x\rightarrow \infty $. \ To do this, there are clearly $n-1$ equivalent
choices. \ We elect to integrate $\frac{\partial }{\partial q_{i_{2}}}$ by
parts to obtain%
\begin{eqnarray*}
&&U_{1}\left( \mathbf{x},t,\mathbf{a}\right) \\
&=&-\varepsilon _{i_{2}\cdots i_{n}}\int \cdots \int dq_{2}\cdots
dq_{n}\,\varepsilon \left( q_{2}-x_{2}\right) e^{a_{2}u_{2}}\frac{\partial }{%
\partial q_{i_{2}}}\left( \frac{e^{a_{1}u_{1}}-1}{a_{1}}\right) \frac{%
\partial }{\partial q_{i_{3}}}\left( \varepsilon \left( q_{3}-x_{3}\right)
e^{a_{3}u_{3}}\right) \cdots \frac{\partial }{\partial q_{i_{n}}}\left(
\varepsilon \left( q_{n}-x_{n}\right) e^{a_{n}u_{n}}\right) \\
&=&-\varepsilon _{i_{2}\cdots i_{n}}\int \cdots \int dq_{2}\cdots
dq_{n}\,\times \\
&&\times \varepsilon \left( q_{2}-x_{2}\right) \frac{\partial u_{1}}{%
\partial q_{i_{2}}}\left( \delta _{i_{3}3}\delta \left( q_{3}-x_{3}\right)
+a_{3}\varepsilon \left( q_{3}-x_{3}\right) \frac{\partial u_{3}}{\partial
q_{i_{3}}}\right) \cdots \left( \delta _{i_{n}n}\delta \left(
q_{n}-x_{n}\right) +a_{n}\varepsilon \left( q_{n}-x_{n}\right) \frac{%
\partial u_{n}}{\partial q_{i_{n}}}\right) \,e^{\mathbf{a\cdot u}}
\end{eqnarray*}%
Expanding out the products of the various paired terms in parentheses in the
last line gives%
\begin{multline*}
U_{1}\left( \mathbf{x},t,\mathbf{a}\right) =-\varepsilon _{i_{2}\cdots
i_{n}}a_{3}\cdots a_{n}\int \cdots \int dq_{2}dq_{3}\cdots
dq_{n}\,\varepsilon \left( q_{2}-x_{2}\right) \,\varepsilon \left(
q_{3}-x_{3}\right) \cdots \varepsilon \left( q_{n}-x_{n}\right) \,\frac{%
\partial u_{1}}{\partial q_{i_{2}}}\frac{\partial u_{3}}{\partial q_{i_{3}}}%
\cdots \frac{\partial u_{n}}{\partial q_{i_{n}}}\,e^{\mathbf{a\cdot u}} \\
-\varepsilon _{i_{2}\cdots i_{n}}\sum_{j=3}^{n}\int dq_{2}\,\varepsilon
\left( q_{2}-x_{2}\right) \,\delta _{ji_{j}}\left( \prod_{\substack{ k=3  \\ %
k\neq j}}^{n}\left( a_{k}\int dq_{k}\,\varepsilon \left( q_{k}-x_{k}\right) 
\frac{\partial u_{k}}{\partial q_{i_{k}}}\right) \right) \frac{\partial u_{1}%
}{\partial q_{i_{2}}}\,e^{\mathbf{a\cdot u}} \\
-\varepsilon _{i_{2}\cdots i_{n}}\sum_{j=3}^{n}\sum_{\substack{ k=4  \\ k>j}}%
^{n}\int dq_{2}\,\varepsilon \left( q_{2}-x_{2}\right) \,\delta
_{ji_{j}}\,\delta _{ki_{k}}\left( \prod_{\substack{ m=3  \\ m\neq j,k}}%
^{n}\left( a_{m}\int dq_{m}\,\varepsilon \left( q_{m}-x_{m}\right) \frac{%
\partial u_{m}}{\partial q_{i_{m}}}\right) \right) \frac{\partial u_{1}}{%
\partial q_{i_{2}}}\,e^{\mathbf{a\cdot u}} \\
-\cdots -\sum_{j=3}^{n}a_{j}\int dq_{2}\,\varepsilon \left(
q_{2}-x_{2}\right) \,\int dq_{j}\,\varepsilon \left( q_{j}-x_{j}\right)
\,\left( \frac{\partial u_{1}}{\partial q_{_{2}}}\frac{\partial u_{j}}{%
\partial q_{_{j}}}-\frac{\partial u_{1}}{\partial q_{_{j}}}\frac{\partial
u_{j}}{\partial q_{2}}\right) \,e^{\mathbf{a\cdot u}} \\
-\int dq_{2}\,\varepsilon \left( q_{2}-x_{2}\right) \,\frac{\partial u_{1}}{%
\partial q_{2}}\,e^{\mathbf{a\cdot u}}
\end{multline*}%
That is to say, the result is given in terms of Nambu brackets \cite{Nambu}
of all ranks from $n-1$ down to $2$ (i.e. Poisson), as well as a final
single derivative term. \ Thus%
\begin{multline*}
U_{1}\left( \mathbf{x},t,\mathbf{a}\right) =-a_{3}\cdots a_{n}\int \cdots
\int dq_{2}dq_{3}\cdots dq_{n}\,\varepsilon \left( q_{2}-x_{2}\right)
\,\varepsilon \left( q_{3}-x_{3}\right) \cdots \varepsilon \left(
q_{n}-x_{n}\right) \,\left\{ u_{1},u_{3},\cdots ,u_{n}\right\} _{23\cdots
n}\,e^{\mathbf{a\cdot u}} \\
-\sum_{j=3}^{n}\int dq_{2}\,\varepsilon \left( q_{2}-x_{2}\right) \left(
\prod_{\substack{ k=3  \\ k\neq j}}^{n}\left( a_{k}\int dq_{k}\,\varepsilon
\left( q_{k}-x_{k}\right) \right) \right) \,\left\{ u_{1},u_{3},\cdots
u_{j-1},u_{j+1},\cdots ,u_{n}\right\} _{23\cdots j-1j+1\cdots n}\,e^{\mathbf{%
a\cdot u}} \\
-\cdots -\sum_{j=3}^{n}\sum_{\substack{ k=4  \\ k>j}}^{n}a_{j}a_{k}\int
dq_{2}\,\varepsilon \left( q_{2}-x_{2}\right) \,\int dq_{j}\,\varepsilon
\left( q_{j}-x_{j}\right) \,\int dq_{k}\,\varepsilon \left(
q_{k}-x_{k}\right) \,\left\{ u_{1},u_{j},u_{k}\right\} _{2jk}\,e^{\mathbf{%
a\cdot u}} \\
-\sum_{j=3}^{n}a_{j}\int dq_{2}\,\varepsilon \left( q_{2}-x_{2}\right)
\,\int dq_{j}\,\varepsilon \left( q_{j}-x_{j}\right) \,\left\{
u_{1},u_{j}\right\} _{2j}\,e^{\mathbf{a\cdot u}}-\int dq_{2}\,\varepsilon
\left( q_{2}-x_{2}\right) \,\frac{\partial u_{1}}{\partial q_{2}}\,e^{%
\mathbf{a\cdot u}}
\end{multline*}%
In the preceding equation, it is to be understood that the sum in the second
RHS row begins at its lower limit with

\noindent $-a_{4}\cdots a_{n}\int \cdots \int dq_{2}dq_{4}\cdots
dq_{n}\,\varepsilon \left( q_{2}-x_{2}\right) \,\varepsilon \left(
q_{4}-x_{4}\right) \cdots \varepsilon \left( q_{n}-x_{n}\right) \,\left\{
u_{1},u_{4},\cdots ,u_{n}\right\} _{24\cdots n}\,e^{\mathbf{a\cdot u}}$

\noindent and terminates at its upper limit with

\noindent $-a_{3}\cdots a_{n-1}\int \cdots \int dq_{2}dq_{3}\cdots
dq_{n-1}\,\varepsilon \left( q_{2}-x_{2}\right) \,\varepsilon \left(
q_{3}-x_{3}\right) \cdots \varepsilon \left( q_{n-1}-x_{n-1}\right)
\,\left\{ u_{1},u_{3},\cdots ,u_{n-1}\right\} _{23\cdots n-1}\,e^{\mathbf{%
a\cdot u}}$. \ 

Next we act with the heat operator on $U_{1}\left( \mathbf{x},t,\mathbf{a}%
\right) $. \ The $\varepsilon $'s permit the appropriate ``outside'' (i.e. $%
x $) partials to be converted, through integration by parts, into ``inside''
(i.e. $q$) partials. \ Also, factors of $a_{i}$ outside the exponentials
produce some extra terms from the cross-partials $\frac{\partial ^{2}}{%
\partial x_{i}\partial a_{i}}$\ in $\mathcal{H}_{n}$. \ We obtain 
\begin{multline*}
\mathcal{H}_{n}U_{1}\left( \mathbf{x},t,\mathbf{a}\right) =-a_{3}\cdots
a_{n}\int \cdots \int dq_{2}dq_{3}\cdots dq_{n}\,\varepsilon \left(
q_{2}-x_{2}\right) \,\varepsilon \left( q_{3}-x_{3}\right) \cdots
\varepsilon \left( q_{n}-x_{n}\right) \,\mathcal{H}_{n}\left( \left\{
u_{1},u_{3},\cdots ,u_{n}\right\} _{23\cdots n}\,e^{\mathbf{a\cdot u}}\right)
\\
+\sum_{i=3}^{n}\frac{\partial }{\partial a_{i}}\left( a_{3}\cdots
a_{n}\right) \int \cdots \int dq_{2}dq_{3}\cdots dq_{n}\,\frac{\partial }{%
\partial x_{i}}\left[ \varepsilon \left( q_{2}-x_{2}\right) \,\varepsilon
\left( q_{3}-x_{3}\right) \cdots \varepsilon \left( q_{n}-x_{n}\right) %
\right] \left\{ u_{1},u_{3},\cdots ,u_{n}\right\} _{23\cdots n}\,e^{\mathbf{%
a\cdot u}} \\
-\sum_{j=3}^{n}\int dq_{2}\,\varepsilon \left( q_{2}-x_{2}\right) \left(
\prod_{\substack{ k=3  \\ k\neq j}}^{n}\left( a_{k}\int dq_{k}\,\varepsilon
\left( q_{k}-x_{k}\right) \right) \right) \,\mathcal{H}_{n}\left( \left\{
u_{1},u_{3},\cdots u_{j-1},u_{j+1},\cdots ,u_{n}\right\} _{23\cdots
j-1j+1\cdots n}\,e^{\mathbf{a\cdot u}}\right) \\
+\sum_{j=3}^{n}\int dq_{2}\,\varepsilon \left( q_{2}-x_{2}\right)
\sum_{i=3}^{n}\frac{\partial }{\partial a_{i}}\frac{\partial }{\partial x_{i}%
}\left( \prod_{\substack{ k=3  \\ k\neq j}}^{n}\left( a_{k}\int
dq_{k}\,\varepsilon \left( q_{k}-x_{k}\right) \right) \right) \,\left\{
u_{1},u_{3},\cdots u_{j-1},u_{j+1},\cdots ,u_{n}\right\} _{23\cdots
j-1j+1\cdots n}\,e^{\mathbf{a\cdot u}} \\
-+\cdots -\sum_{j=3}^{n}a_{j}\int dq_{2}\,\varepsilon \left(
q_{2}-x_{2}\right) \,\int dq_{j}\,\varepsilon \left( q_{j}-x_{j}\right) \,%
\mathcal{H}_{n}\left( \left\{ u_{1},u_{j}\right\} _{2j}\,e^{\mathbf{a\cdot u}%
}\right) \\
+\sum_{i=3}^{n}\frac{\partial }{\partial a_{i}}\frac{\partial }{\partial
x_{i}}\left( \sum_{j=3}^{n}a_{j}\int dq_{2}\,\varepsilon \left(
q_{2}-x_{2}\right) \,\int dq_{j}\,\varepsilon \left( q_{j}-x_{j}\right)
\right) \,\left\{ u_{1},u_{j}\right\} _{2j}\,e^{\mathbf{a\cdot u}} \\
-\int dq_{2}\,\varepsilon \left( q_{2}-x_{2}\right) \,\mathcal{H}_{n}\left( 
\frac{\partial u_{1}}{\partial q_{2}}\,e^{\mathbf{a\cdot u}}\right)
\end{multline*}%
The first RHS line of $\mathcal{H}_{n}U_{1}$ reduces to terms linear in the
E-M equations for the $u$'s. \ The second and third RHS lines combine to
give similar terms linear in the E-M equations. \ And so it goes with
subsequent pairs of RHS lines, until finally the last two RHS lines combine
to give terms linear in the E-M equations. \ 

To establish these statements, one needs to use several identities involving
the action of the heat operator on exponentially weighted derivatives of the
component fields, in particular on so-weighted Nambu brackets. \ For
example, these identities range from the simplest for the full Jacobian%
\begin{multline*}
\mathcal{H}_{n}\left( e^{\mathbf{a\cdot u}}\left\{ u_{1},u_{2},\cdots
,u_{n}\right\} _{12\cdots n}\right) =e^{\mathbf{a\cdot u}}\left( \mathbf{%
a\cdot }\mathcal{M}_{n}\mathbf{u}\right) \left\{ u_{1},u_{2},\cdots
,u_{n}\right\} _{12\cdots n} \\
+e^{\mathbf{a\cdot u}}\left( \left\{ \mathcal{M}_{n}u_{1},u_{2},\cdots
,u_{n}\right\} _{12\cdots n}+\left\{ u_{1},\mathcal{M}_{n}u_{2},\cdots
,u_{n}\right\} _{12\cdots n}+\cdots +\left\{ u_{1},u_{2},\cdots ,\mathcal{M}%
_{n}u_{n}\right\} _{12\cdots n}\right)
\end{multline*}%
to those involving lower rank Nambu brackets such as%
\begin{multline*}
\mathcal{H}_{n}\left( e^{\mathbf{a\cdot u}}\left\{ u_{2},u_{3},\cdots
,u_{n}\right\} _{23\cdots n}\right) =e^{\mathbf{a\cdot u}}\left( -\left\{
u_{1},u_{2},u_{3},\cdots ,u_{n}\right\} _{123\cdots n}+\left( \mathbf{a\cdot 
}\mathcal{M}_{n}\mathbf{u}\right) \left\{ u_{2},u_{3},\cdots ,u_{n}\right\}
_{23\cdots n}\right) \\
+e^{\mathbf{a\cdot u}}\left( \left\{ \mathcal{M}_{n}u_{2},u_{3},\cdots
,u_{n}\right\} _{23\cdots n}+\left\{ u_{2},\mathcal{M}_{n}u_{3},\cdots
,u_{n}\right\} _{23\cdots n}+\cdots +\left\{ u_{2},u_{3},\cdots ,\mathcal{M}%
_{n}u_{n}\right\} _{23\cdots n}\right)
\end{multline*}%
including that needed to deal with the first RHS line of $\mathcal{H}%
_{n}U_{1}$%
\begin{multline*}
\mathcal{H}_{n}\left( e^{\mathbf{a\cdot u}}\left\{ u_{1},u_{3},\cdots
,u_{n}\right\} _{23\cdots n}\right) =e^{\mathbf{a\cdot u}}\left( \mathbf{%
a\cdot }\mathcal{M}_{n}\mathbf{u}\right) \left\{ u_{1},u_{3},\cdots
,u_{n}\right\} _{23\cdots n} \\
+e^{\mathbf{a\cdot u}}\left( \left\{ \mathcal{M}_{n}u_{1},u_{3},\cdots
,u_{n}\right\} _{23\cdots n}+\left\{ u_{1},\mathcal{M}_{n}u_{3},\cdots
,u_{n}\right\} _{23\cdots n}+\cdots +\left\{ u_{1},u_{3},\cdots ,\mathcal{M}%
_{n}u_{n}\right\} _{23\cdots n}\right)
\end{multline*}%
as well as other relations obtained by permutations of the indices of these,
etc., all the way down to the final%
\begin{equation*}
\mathcal{H}_{n}\left( e^{\mathbf{a\cdot u}}\partial _{j}u_{k}\left( \mathbf{x%
},t\right) \right) =e^{\mathbf{a\cdot u}}\left( \partial _{j}\left( \mathcal{%
M}_{n}u_{k}\right) +\left( \mathbf{a\cdot }\mathcal{M}_{n}\mathbf{u}\right)
\partial _{j}u_{k}-\sum_{i}\left\{ u_{k},u_{i}\right\} _{ji}\right)
\end{equation*}%
as needed to deal with the last two RHS lines in $\mathcal{H}_{n}U_{1}$. \
All such identities are straightforward to substantiate by direct
calculation.

Thus, given the E-M equations for the $u$'s, the heat equation for $U_{1}$
follows. \ Moreover, the only terms on the RHS of $\mathcal{H}_{n}U_{1}$
which survive in the limit of vanishing spectral parameters are the last two
lines, which give%
\begin{equation*}
\lim_{\mathbf{a}\rightarrow \mathbf{0}}\mathcal{H}_{n}U_{1}\left( \mathbf{x}%
,t,\mathbf{a}\right) =\mathcal{M}_{n}u_{1}\left( \mathbf{x},t\right)
\end{equation*}%
Thus, given the heat equation for $U_{1}$, the E-M equation for $u_{1}$
follows. \ Similar results obtain for all the other components, so that $%
\mathcal{H}_{n}U_{k}=0$ iff $\mathcal{M}_{n}u_{j}=0$. \ $\blacksquare $

\paragraph{Corollary I of {\protect\large n} in {\protect\large n}:}

Formally, time evolution in the bulk\ is given by%
\begin{equation*}
\mathbf{U}\left( \mathbf{x},t,\mathbf{a}\right) =e^{t\sum_{j=1}^{n}\frac{%
\partial ^{2}}{\partial x_{j}\partial a_{j}}}\mathbf{U}\left( \mathbf{x},t=0,%
\mathbf{a}\right)
\end{equation*}%
This gives a time-series solution on the boundary upon taking the limit $%
\mathbf{a}\rightarrow \mathbf{0}$.%
\begin{equation*}
\mathbf{u}\left( \mathbf{x},t\right) =\lim_{\mathbf{a}\rightarrow \mathbf{0}%
}e^{t\sum_{j=1}^{n}\frac{\partial ^{2}}{\partial x_{j}\partial a_{j}}}%
\mathbf{U}\left( \mathbf{x},t=0,\mathbf{a}\right)
\end{equation*}%
with initial boundary data $\mathbf{u}\left( \mathbf{x}\right) =\lim\limits_{%
\mathbf{a}\rightarrow \mathbf{0}}\mathbf{U}\left( \mathbf{x},t=0,\mathbf{a}%
\right) $.

\paragraph{Corollary II of {\protect\large n} in {\protect\large n}:}

The n-fold infinite sequences of conservation laws for the E-M equations in $%
n$ spatial dimensions are directly encoded into the bulk solutions.%
\begin{equation*}
\frac{\partial }{\partial t}U_{k}\left( \mathbf{x},t,\mathbf{a}\right)
=\nabla \cdot \mathbf{J}_{k}\left( \mathbf{x},t,\mathbf{a}\right) \;,\;\;\;%
\mathbf{J}_{k}\left( \mathbf{x},t,\mathbf{a}\right) =\nabla _{\mathbf{a}%
}U_{k}\left( \mathbf{x},t,\mathbf{a}\right) \;,\;\;\;k\in \left\{ 1,2,\cdots
,n\right\} \;.
\end{equation*}%
Explicit sequences of charge and current densities on the boundary follow
immediately from power series expansions in the $a_{j}$. \ Both Corollaries
are elementary consequences of the Theorem and the heat equation obeyed by $%
\mathbf{U}$. \ $\blacksquare $

This is as far as we have completed the application of the higher
dimensional approach to classical nonlinear PDEs. \ It remains to apply this
approach to other types of nonlinear PDEs, in particular to those
higher-order extensions of the E-M equations involving dispersion, such as
the Korteweg-deVries equation, and to those involving diffusion, such as the
Burgers and Navier-Stokes equations. \ Another immediately obvious challenge
is to carry the method over to quantum field theories (QFTs). \ This will
not be done here. \ However, we suspect that the implementation of these
ideas in QFT will involve the use of quantum Nambu brackets (QNBs), given
that the classical versions of these appear above. \ QNBs have a
long-standing notoriety, but recently \cite{CurtrightZachos} it has been
shown that theirs is an undeserved bad reputation. \ QNBs can be defined in
terms of operators (or in terms of non-commutative geometry) so as to fulfil
their expected roles in the quantum evolution of dynamical systems. \
Perhaps these developments will be useful to meet the challenge of
quantizing the E-M equations as well as their higher-order generalizations.

As emphasized previously, the Euler-Monge equations appear widespread
throughout physics and the mathematics of nonlinear partial differential
equations. \ Based on the maps we have presented to linearize these
equations, we have come to the following remarkable conclusion. \ Extra
dimensions and nonlocal structures are probably universally applicable
features to be found upon analyzing solutions of nonlinear partial
differential equations, and hence quite natural constructs in almost all
physical theories.

\paragraph{Acknowledgements:}

Our thanks to Cosmas Zachos for valuable comments, and to Orlando Alvarez
for critically reading the manuscript. \ This research was supported in part
by a Leverhulme Emeritus Fellowship and by NSF Award 0073390. \


\begin{thebibliography}{99}
\bibitem{Appelquist} T Appelquist, A Chodos, and P G O Freund, \textit{%
Modern Kaluza-Klein Theories}, Addison-Wesley, 1987.

\bibitem{Arkani-Hamed} N Arkani-Hamed, A G Cohen, and H Georgi,
``(De)Constructing Dimensions'', Phys Rev Lett \textbf{86} (2001) 4757-4761;
\ C T Hill and A K Leibovich, ``Deconstructing 5-D QED'', hep-ph/0205057.

\bibitem{Baker} L Baker and D B Fairlie, ``Hamilton-Jacobi equations and
Brane associated Lagrangians'', Nucl Phys \textbf{B596} (2001) 348-364.

\bibitem{Bateman} H Bateman, ``Some recent researches on the motion of
fluids'', Monthly Weather Rev. \textbf{43} (1915) 163-170.

\bibitem{Burgers} J M Burgers, \textit{The Nonlinear Diffusion Equation},
Reidel, Dordrecht, 1974.

\bibitem{ColeHopf} J D Cole, ``On a quasi-linear parabolic equation
occurring in aerodynamics'', Quart Appl Math \textbf{9} (1951) 225-236; \ E.
Hopf, ``The partial differential equation $u_{t}+u\,u_{x}=\mu _{xx}$'', Comm
Pure Appl Math \textbf{3} (1950) 201-230.

\bibitem{CurtrightZachos} T Curtright and C Zachos, ``Deformation
Quantization of Superintegrable Systems and Nambu Mechanics'',
hep-th/0205063.

\bibitem{Debnath} L Debnath, \textit{Nonlinear partial differential
equations for scientists and engineers}, Birkh\"{a}user, 1997.

\bibitem{Dubrovin} B A Dubrovin and S P Novikov, ``Hydrodynamics of weakly
deformed soliton lattices. \ Differential geometry and Hamiltonian theory'',
Russian Math. Surveys \textbf{44} (1989) 35-124.

\bibitem{Euler} L Euler, \textit{Opera Omnia}, Series II, \textit{Opera
mechanica et astronomica}, Volumes 12 and 13, \textit{Commentationes
mechanicae ad theoriam corporum fluidorum pertinentes}, edited by Clifford
Ambrose Truesdell, Birkh\"{a}user.

\bibitem{dbf1} D B Fairlie, J Govaerts and A Morozov, ``Universal Field
Equations with Covariant Solutions'', Nucl Phys \textbf{B373} (1992) 214-232.

\bibitem{dbf2} D B Fairlie, ``Integrable Systems in Higher Dimensions'' 
\textit{Quantum Field Theory, Integrable Models and Beyond}, Editors T.
Inami and R. Sasaki , Progress of Theoretical Physics Supplement \textbf{118}
(1995) 309-327.

\bibitem{dbf3} D B Fairlie, ``Formal Solutions of an Evolution Equation of
Riemann type'', Studies in Applied Math \textbf{98} (1997) 203-205.

\bibitem{dbf4} D B Fairlie, ``A Universal Solution'', accepted for
publication in J Nonlin Math Phys, math-ph/0112033.

\bibitem{Forsyth} A R Forsyth, \textit{Theory of Differential Equations.
Part IV -- Partial Differential Equations}, Cambridge University Press,
Cambridge, 1906. Republished by Dover, New York, 1959. \ In particular, see
page 100.

\bibitem{Gesztesy} F Gesztesy and H Holden, ``The Cole-Hopf and Miura
Transformations Revisited'', solv-int/9812025.

\bibitem{Monge} G Monge, \textit{Application de l'analyse a la g\'{e}om\'{e}%
trie}, Bernard, 1807 (H. Perronneau).

\bibitem{Nambu} Y Nambu, ``Generalized Hamiltonian Dynamics'', Phys Rev 
\textbf{D7} (1973) 2405-2412.

\bibitem{Polyakov} A Polyakov, ``Turbulence without pressure'', Phys Rev 
\textbf{E52} (1995) 6183-6188.

\bibitem{Riemann} B Riemann ``Uber die Fortpflanzung ebener Luftwellen von
endlicher Schwingungsweite'', G\"{o}ttingen Abhandlunger, Vol. viii, p 43 (%
\textit{Werke}, 2te Aufl., Leipzig, 1892, p 157).

\bibitem{Whitham} G B Whitham, \textit{Linear and Nonlinear Waves}, Wiley,
1974.
\end{thebibliography}
\end{document}